# Pre-perihelion photometric behavior of comet C/2012 S1 (ISON) and its future prospect

Jakub Cerny

*Society for interplanetary matter, Kraví hora 522/2, Brno-střed, 616 00, Czech Republic*
*Version November 11, 2013*

**Abstract**

*Comet C/2012 S1 (ISON) shows strange photometric behavior compared to many past comets. Shortly after its discovery, majority of astronomers expected its surviving until perihelion passage. Unprecedented evolution of its photometric parameters nearly 2 AU from Sun, pushed this comet under survival line defined by formula of John. E. Bortle, where ~70% comets in his analysis disintegrated. However comparison with photometric behavior of group of past disintegrated comets showing large differences and ISON doesn't seems to be typical member of this group. In other hand, compared to surviving, dynamically new comets which were observed far away from Sun and survived its perihelion passage, ISON showing peculiar gradual decrease of its activity, probably caused by exhaustion of active fractions of its nucleus.*

## 1. Introduction

In year 1991 John E. Bortle made simply comparison of absolute magnitudes $H_{10}$ from sample of 85 comets - 4 short periodic, 81 long periodic, from which 16 didn't survive its perihelion passage. He find out that below line defined by formula $H_{10} = 7.0 + 6q$ (where q is perihelion distance in AU), nearly 70% of comets disintegrated [1]. The resulting value for perihelion distance 0.0124464 AU [2] for comet C/2012 S1 (ISON) is 7.07 mag which is far brighter then actual brightness of this comet nearly 1 AU from Sun. Therefore this comet can be in potential danger of its nucleus disintegration before reaching perihelion or shortly after it.

Earlier observations in larger distances from Sun suggested originally intrinsically much brighter object, but later unprecedented gradual decline of its cometary activity dropped it below this line. Photometric behavior of this comet in larger heliocentric distance can play key role in predicting its fate near perihelion. Also comparison with disintegrated comets may help finding more common indicators for this group except $H_{10}$ value.

Sample of surviving comets was selected by their original orbit and length of their pre-perihelion follow up. Only dynamically new comets, which are believed to be first time in inner part of Solar system was selected, with perihelion distance smaller then 1 AU and observed is larger then 3 AU distances pre-perihelion. From total number of 7 comets all of them have original distance of semi-major axis larger then 20 000 AU from Sun. Only exception is comet C/2000 WM1 (LINEAR) with original semi-major axis 1878 [3], however it was included to study because of some similar photometric behavior with comet ISON.

Sample of disintegrated comets is far more complicated and because of very poor orbit determination it is unclear which can be true dynamically new comets. Also some of selected comets doesn't have its disintegration confirmed. C/2009 R1 (McNaught) was observed only before perihelion passage, after it almost all attempts to picture this comet failed and only one unconfirmed positive observation was reported. Intrinsically extremely faint comet C/2012 T5 (Bressi) was followed before reaching perihelion, and after this undergone strong outburst, reported by David Sergeant 4.44 February 2013 [4]. It faded after this and only 3 measures of its astrometric position was done post-perihelion in only one night by observatory with MPC code 958 [5], without any other confirming positive reports.

## 2. Processing of photometric data

Photometry of comets is very difficult discipline because of many aspects. Both visual and CCD measures suffering large scatter caused by observing conditions, used methods and different equipment. Data sets for this study was selected from COBS database [6], Astrosite Groningen [7] and MPC database [8], because they allow easy processing of large data sets. To avoid false results of photometric behavior and minimize differences between selected data sets, some corrections are used. Because of compatibility with original study of John. E. Bortle, visual data sets are always used as reference. They are also used only in intervals, where additional CCD data doesn't provide any significant improvement in light curve precision to avoid redundant corrections.

Correction coefficients of various CCD data sets are determined by comparing with visual magnitude sets in interval with presence of both sources in pre-perihelion time, post-perihelion have no use for this study. In selected interval, best fit photometric parameters are calculated for one (with more data points) data set and resulting brightening slope factor n is used to calculate $H_{10}$ parameter for other set, CCD magnitudes are then corrected by resulting $H_{10}$ difference in whole range. Such correction is needed to avoid artificial change of slope caused by transition from CCD to visual observations when comets reach range of visual observers.

From MPC data sets, only brightest total magnitudes in scatter up to ~1 mag in selected intervals with length of one month (or smaller if real magnitude changes faster) are used, because of extremely large scatter of this data source. If dramatically large incompatibility with regular CCD photometry and visual data sets is found, more data are excluded from final analysis.

It is necessary to note all possible errors. Because color index of comets is changing in time with different contributions of reflected sunlight on dust and gas molecules emission lines to total magnitude with different distance from Sun. Systematical errors caused by changes in real coma diameter and using bad apertures to measure total CCD magnitudes can also occurs. Applying corrections for both effects in unfortunately not possible, because of poor and unclear coverage of them for all comets. Correction for phase angle was also not used. However for fast analysis and prediction needs in near future time, resulting data sets quality seems to be sufficient for this purpose.

## 3. Light curve analysis

First analysis was done for comet C/2012 (ISON) and possible evolution of its $H_{10}$ parameter in various heliocentric distances. Assuming fixed slope parameter n = 4 in equation $H_{10} = m1 - 5 \log D - 2.5n \log r$, resulting values for different heliocentric distance with step 1 AU are over 1.5 higher then its survival limit for wide range of r between 9 AU to 5 AU. At r = 4 AU, $H_{10}$ value started to gradually dropping after r = 4 AU and near r = 2 AU it fall under survival limit. Very dramatic decline happened between r 2 and 1 AU, when $H_{10}$ dropped more then 1.5 mag under survival limit.

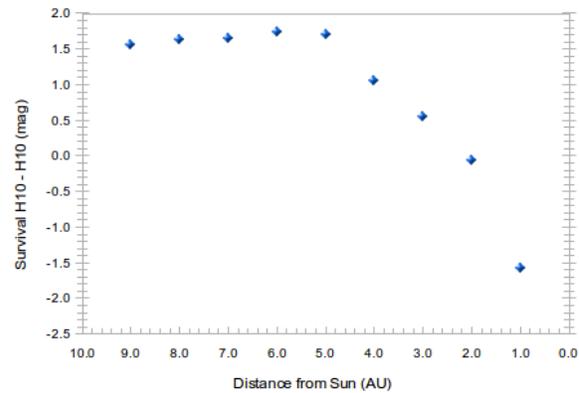

**Figure 1.** Plot of difference between survival limit of comet C/2012 S1 (ISON) and $H_{10}$ values calculated for different heliocentric distances using fixed slope parameter n = 4.

Such evolution may suggest, that comet C/2012 S1 (ISON) have nucleus large enough to be over surviving limit and past sudden decrease in activity closer to Sun is just transient phenomenon. Unfortunately no comparison with extinct comets can be done, because

**Table 1**

Difference in brightness change (mag) of C/2012 S1 (ISON) between various r (AU) against comparison comets.

| Comet | 9 –> 1 | 8 –> 1 | 7 –> 1 | 6 –> 1 | 5 –> 1 | 4 –> 1 | 3 –> 1 | 2 –> 1 |
|---|---|---|---|---|---|---|---|---|
| C/2001 Q4 (NEAT) | -0.92 | -1.13 | -1.38 | -1.62 | -1.47 | -0.71 | -0.30 | +0.15 |
| C/2011 L4 (PANSTARRS) | | -3.62 | -3.09 | -2.53 | -1.67 | -0.78 | -0.06 | -0.23 |
| C/2002 T7 (LINEAR) | | | | -2.45 | -2.39 | -1.74 | -1.25 | -0.57 |
| C/2003 K4 (LINEAR) | | | | -1.53 | -1.38 | -1.04 | -0.71 | +0.23 |
| C/2000 WM1 (LINEAR) | | | | | -0.92 | -0.98 | -1.25 | -0.86 |
| C/2007 W1 (Boattini) | | | | | | -1.66 | -2.14 | -2.48 |
| C/2006 P1 (McNaught) | | | | | | | -5.09 | -2.54 |
| **Average** | **-0.92** | **-2.37** | **-2.23** | **-2.04** | **-1.57** | **-1.15** | **-1.54** | **-0.90** |
| **Median** | **-0.92** | **-2.37** | **-2.23** | **-2.04** | **-1.47** | **-1.01** | **-1.25** | **-0.57** |

they were never been observed in such large heliocentric distances and it is unknown if they can exhibit such activity decrease while approaching Sun, or their activity levels are lower permanently.

### 3.1. Surviving dynamically new comets

What can be done is comparison with other dynamically new comets, observed far away from Sun. Assumption that composition and active fraction of their nuclei is similar amongst their population, even with little differences in water/volatiles composition and dust-to-gas ratio should lead to similar brightness evolution while approaching to Sun. Therefore difference in brightness change between various heliocentric distances should be similar amongst population of this comets.

Results from analysis in table 1 shows that brightness growth of comet ISON is lagging behind all comets from selected sample while approaching to Sun except two cases between $r = 2$ and $r = 1$. Activity growth slowdown of ISON is unprecedented amongst dynamically new comets that survived. In distances 5 AU and farther, brightness of comet ISON is similar and not fainter then 2 mag then any comparison comet.

**Table 2**

List of analyzed comets and determined original semimajor axes.

| Comet | Orig. semimajor axis | Source |
|---|---|---|
| C/2012 S1 (ISON) | 103 842 | MPEC 2013-V48 |
| **Surviving dynamically new comets** | | |
| C/2000 WM1 (LINEAR) | 1 878 | MPC 46619 |
| C/2001 Q4 (NEAT) | 23 063 | MPC 52163 |
| C/2002 T7 (LINEAR) | 21 057 | MPC 52164 |
| C/2003 K4 (LINEAR) | 42 772 | MPC 52315 |
| C/2006 P1 (McNaught) | 32 404 | MPC 82313 |
| C/2007 W1 (Boattini) | -8 995 | MPC 63599 |
| C/2011 L4 (PANSTARRS) | 33 201 | MPEC 2013-V48 |
| **Disintegrating comets** | | |
| C/1996 Q1 (Tabur) | 548 | MPC 28052 |
| C/1999 S4 (LINEAR) | -18 345 | MPC 40988 |
| C/2000 W1 (Utsunomiya-Jones) | | MPC 42106 |
| C/2002 O4 (Hoenig) | -1 295 | MPC 46762 |
| C/2002 O6 (SWAN) | 344 | [10] |
| C/2002 O7 (LINEAR) | 37 369 | MPC 47292 |
| C/2009 O2 (Catalina) | 2 876 | MPC 69905 |
| C/2009 R1 (McNaught) | -31 447 | MPC 82315 |
| C/2010 X1 (Elenin) | 111 111 | [10] |
| C/2012 T5 (Bressi) | -4 401 | MPEC 2013-J25 |

In distance 1 AU from Sun, ISON is nearly 3 mag fainter then all of them except C/2000 WM1 (LINEAR) and C/2007 W1 (Boattini). Brightening lag is most constant compared to comet C/2000 WM1 (LINEAR), which brightness evolution is only one comparable to ISON.

### 3.2. Disintegrating comets

There were many cases of comets which disintegrates is past years. It is very hard to determine presence of dynamically new population amongst them, because their orbit are usually poorly determined. Reason is usually small observed arc of their orbit and large influence of non-gravitational forces. Comet C/1996 Q1 (Tabur) was already recognized as smaller fragment of split comet together with C/1988 A1 (Liller) [9]. Very small original semi-major axis of comet C/2002 O6 (SWAN) found by Kinoshita [10] pointing to similar nature of this comet.

**Table 3**

Brightening slope parameter n before disintegration.

| Comet | n |
|---|---|
| C/1996 Q1 (Tabur) | 5.48 |
| C/1999 S4 (LINEAR) | 5.22 |
| C/2000 W1 (Utsunomiya-Jones) | 4.15 |
| C/2002 O4 (Hoenig) | 9.31 |
| C/2002 O6 (SWAN) | 5.35 |
| C/2002 O7 (LINEAR) | 2.97 |
| C/2009 O2 (Catalina) | 3.54 |
| C/2009 R1 (McNaught) | 3.94 |
| C/2010 X1 (Elenin) | 4.23 |
| C/2012 T5 (Bressi) | - |

From other 8 comets in Table 2, half exhibit original orbit to be hyperbolic. This can be result of strong non-gravitational forces on small, very fast eroding nucleus described by Sekanina [11]. It is also noticeable, that amongst "health" population, only C/2007 W1 exhibits such feature, intrinsically faintest of sample with exceptionally large non-gravitational forces (A1 = 3.847, A2 = -0.941374 [12]). With perihelion more close to Sun, this comet would be probably in serious threat of potentially disintegration.

Ignacio Ferrin [13] predicted demise of comet ISON before reaching its perihelion based on analysis of light curve shapes determined for few disintegrated comets. He state that early warning signal of disintegration can be flattening of light curve, where magnitude continue grow as simple reflecting body or still.

In this study, other aspects of photometric behavior like relative evolution of activity in similar distances from Sun, was compared. Majority of disintegrated comets exhibits very short and sharp light curve with originally low activity unlike comet ISON. On analyzed light curves is easy to see position of terminal disintegration events. Resulting slope parameters n from last stable interval before terminal disintegration are collected in Table 3. Most of them exceed normal average n = 4 for comets, while all was brightening faster then simple reflecting body (n = 2). Therefore flattening of light curve can't be taken as early warning signal of incoming disintegration, but naturally it is signal that such process already ended and there is no more present nucleus which can fuel continuous growth of cometary activity. Unprecedented continuous fading of comet ISON doesn't seems to be sign of its demise before reaching perihelion.

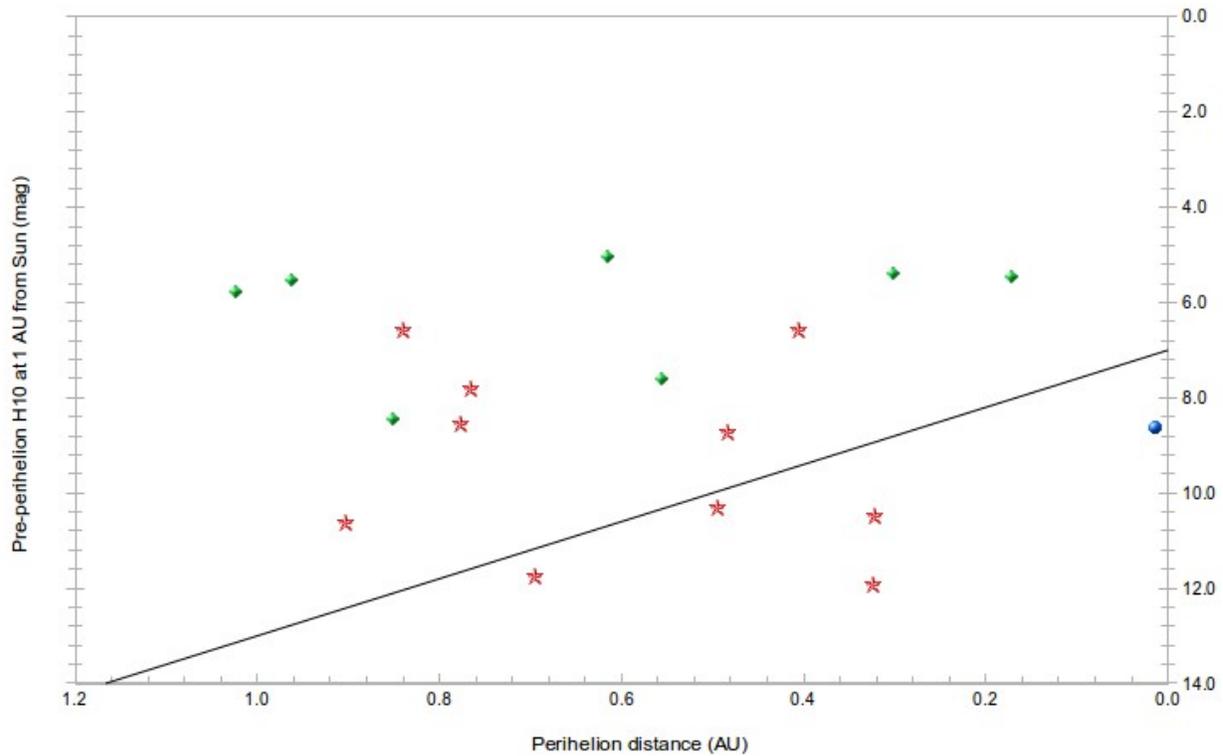

**Figure 2**. $H_{10}$ values for analyzed comets at r = 1 AU pre-perihelion. Surviving, dynamically new comets are diamonds, disintegrated and probably disintegrated comets are stars, circle is comet C/2012 S1 (ISON). Line representing surviving limit defined by Bortle.

From comparison of light curves between extinct comets and ISON it is also noticeable that activity of some extinct comets were close to ISON at least in some parts of its light curve. However it is unclear if this is not just an result of rapid progressive disintegrating process, instead of normal activity level for such comet category. For example, Sekanina stated that dust production peak at the end of disintegration process of comet C/2002 O4 exceeded normal dust production rate of comet 1P/Halley at same heliocentric distance [14]. Figure 2 with $H_{10}$ plotted for all analyzed comets shows that many past disintegrated comets are situated over their surviving limit including comet C/2002 O4, in contrast with original work of Bortle, where there were no disintegrating comets over that line. Such effect can be caused by different nature of objects found in past years. Compared to historical comets, most objects recorded in past years could be unnoticed before 1991 and their disintegration is placed in larger distances from Sun. Close to 1 AU where $H_{10}$ for this plot was calculated, the rapid progressive event could be already taking place and the comets looks temporary brighter because of short-lived extensive dust and gas productions.

## 4. Conclusions

This study shows serious difference between comet ISON and other comets, no matter if they disintegrated or no. Except very low $H_{10}$ value in heliocentric distance 2 AU and smaller, there is no evidence that ISON should demise before reaching its perihelion. This conclusion is supported by its orbital evolution and lack of non-gravitational forces stated in work of Sekanina [11]. Small $H_{10}$ value can be resulting from exhaustion of surface reservoirs of volatiles and water ice on relative large nucleus.

Comparison with sample of others dynamically new comets showing unprecedented fading of originally very active object. In recent study, Jian-Yang Li and his team analyzed orientation of sun-ward jet and its movement in period of time on images from Huble Space Telescope [15]. Most possible result shows that rotational axis of nucleus is pointing to proximity of Sun which means that large portion of nucleus is in constant darkness, while other experiencing permanent heating from Sun. This can explain why large distant activity of comet exhausted so early. HST image from 9. October 2013 shows that strong jet, visible on images from 10. April earlier this year, already disappeared [16].

This theory seems to be consistent with unprecedented photometric behavior of this comet. If this is correct, during its perihelion passage, relative orientation of comet nucleus to Sun should dramatically change and heating of previously darkened surface may cause at least partially returning of previous activity of this comet (it is lacking nearly 2 magnitudes compared with 4 comets at r = 6 AU as its seen in Table 1). It can lead to similar effect observed in case of analyzed comet C/2000 WM1 (LINEAR) which brightened after passing perihelion by nearly 3 magnitudes [17]. However perihelion distance of comet ISON is much smaller then all analyzed comets and surviving perihelion passage itself is not part of this study.

## 6. Figures

Comparison plots between comet C/2012 S1 (ISON) and other comets analyzed in this study is attached. Distance corrected magnitudes are plotted against log(r). Program used for analysis is Comet for windows, programmed by Seiichi Yoshida. ISON observations are always plotted as black points, its light curve is also black. Red points and light curves are for comparison comets.

## 6.1. Surviving dynamically new comets and comet C/2012 S1 (ISON)

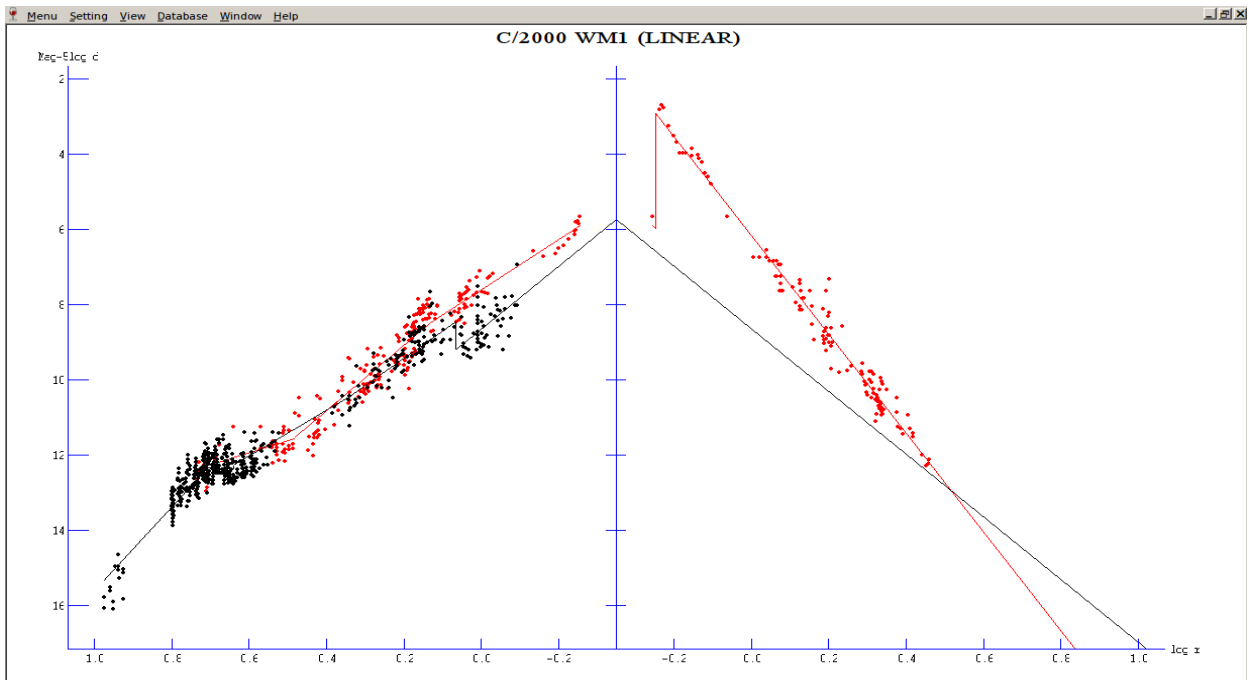

**Figure 3**. Comet C/2000 WM1 (LINEAR) is probably not true dynamically new object near Sun, from all compared comet, evolution of its activity is far most similar to comet ISON.

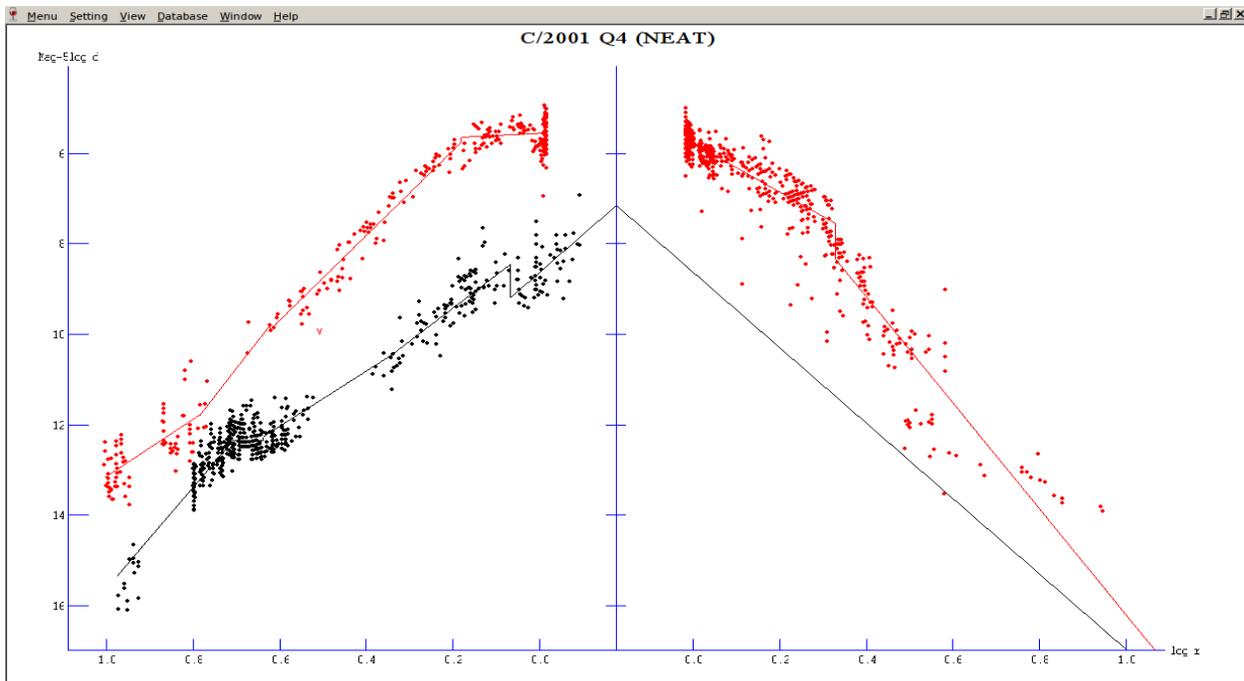

**Figure 4**. Comet C/2001 Q4 (NEAT) exhibit larger total activity then ISON, and exhibited strong flattening of its light curve near Sun.

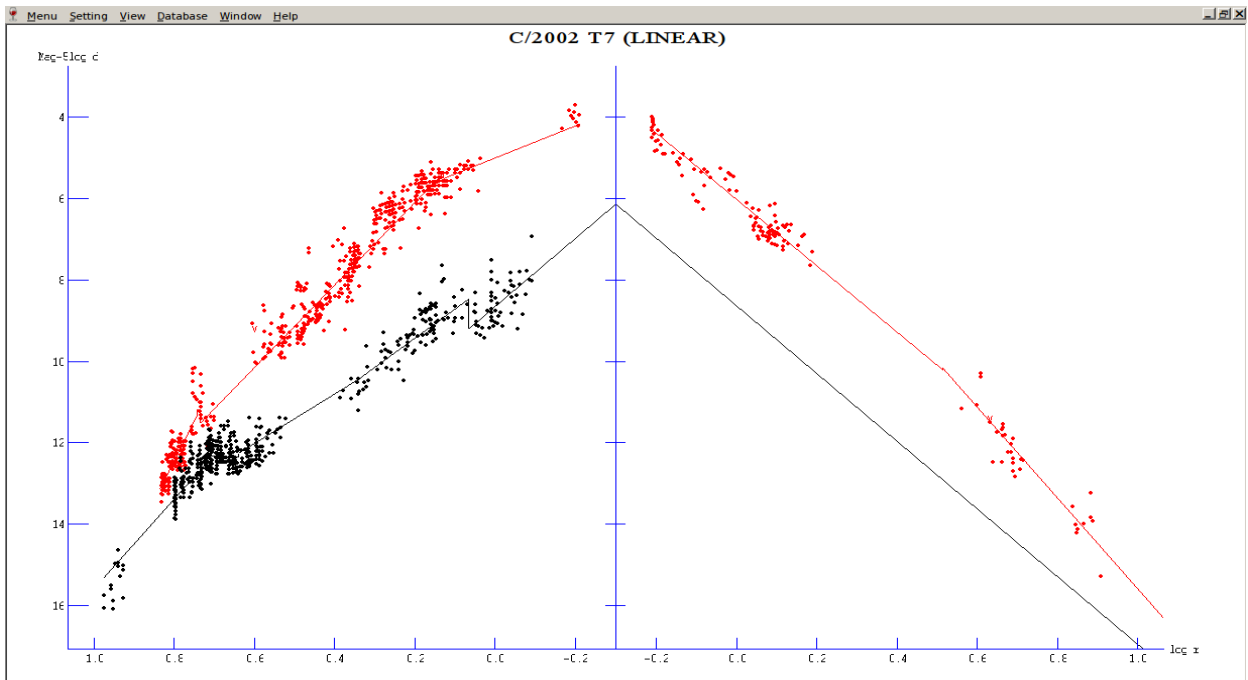

**Figure 5**. Comet C/2002 T7 (LINEAR) was slightly more active then ISON when it was discovered, during it approaching to Sun its activity grow very fast then undergone slowdown of its activity before reaching perihelion.

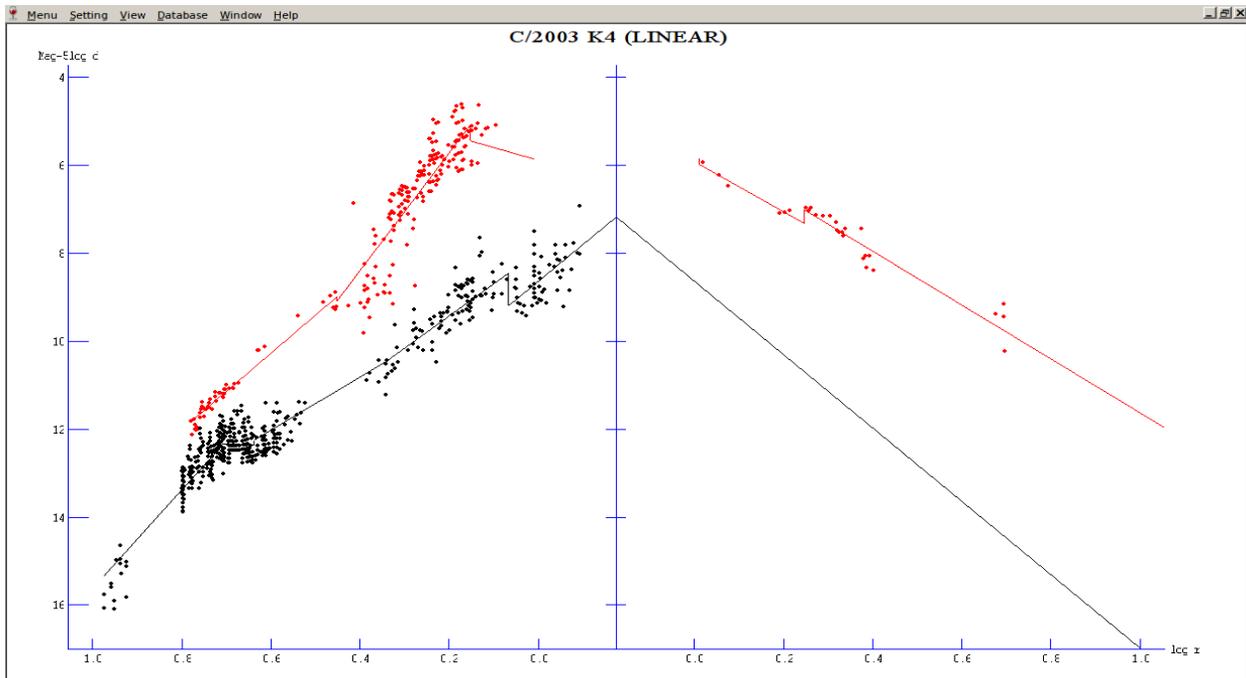

**Figure 6**. Comet C/2003 K4 (LINEAR) exhibited fast growth of its brightness then reached maximal brightness before perihelion, after it it little faded.

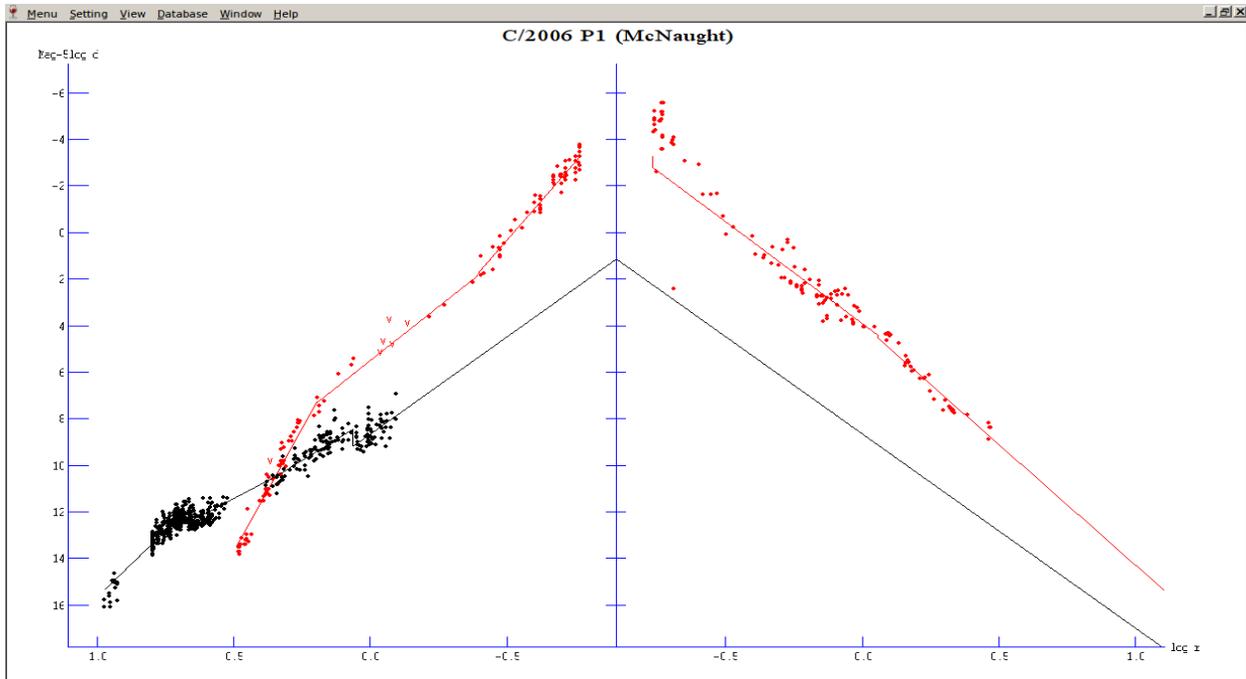

**Figure 7**. Famous comet C/2006 P1 (McNaught) seems to be originally fainter then ISON and originally it was not expected to survive its perihelion passage. However its activity reached very high level and it become a "Great comet" later.

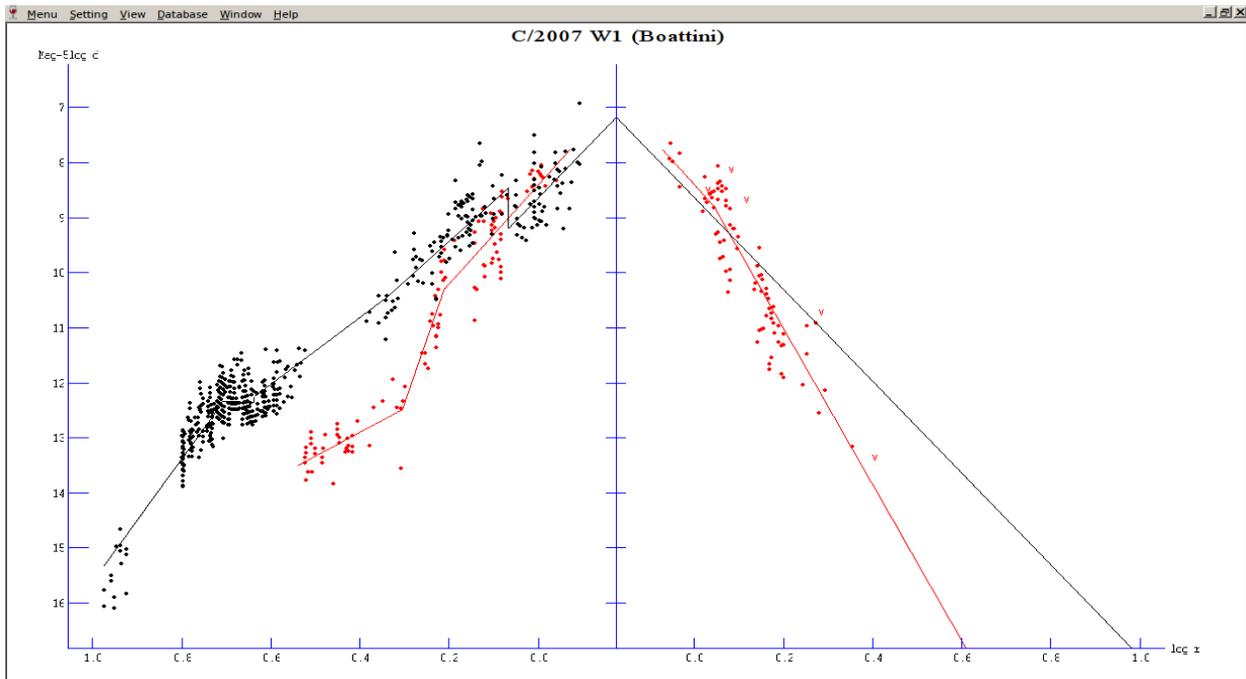

**Figure 8**. C/2007 W1 (Boattini), originally very faint comet with strong non-gravitational could be in danger of disintegration if it approach closer to Sun.

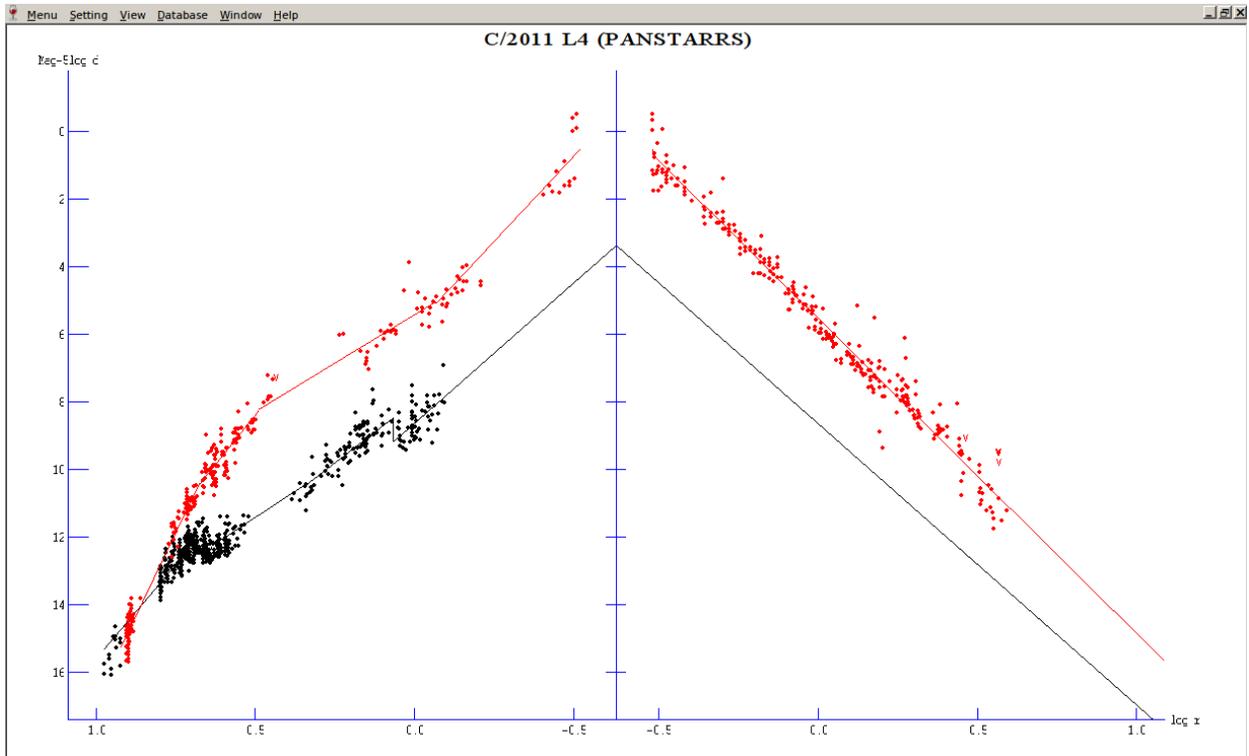

**Figure 9**. C/2011 L4 (PANSTARRS) shows originally extremely similar brightness as comet ISON, somehow its activity later exceeded levels of ISON by almost 4 mag.

## 6.2. Disintegrating comets and comet C/2012 S1 (ISON)

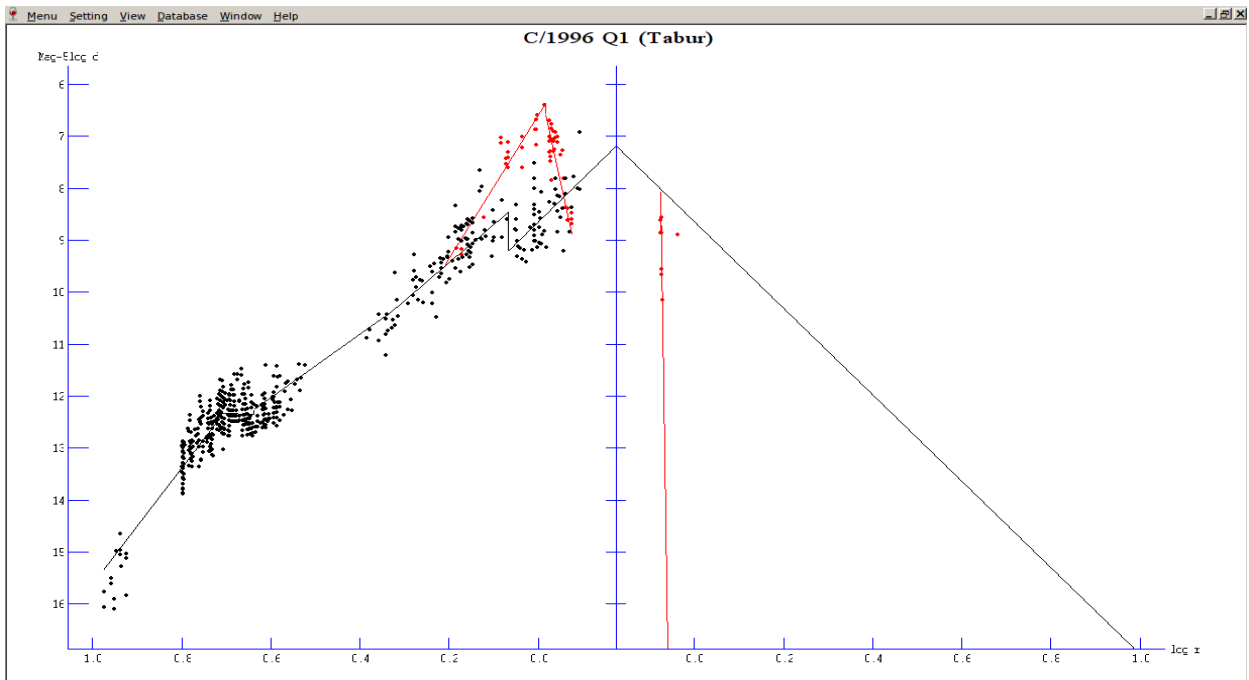

**Figure 10**. C/1996 Q1 (Tabur) was observed only shortly and exhibited very strong peak ended with its demise.

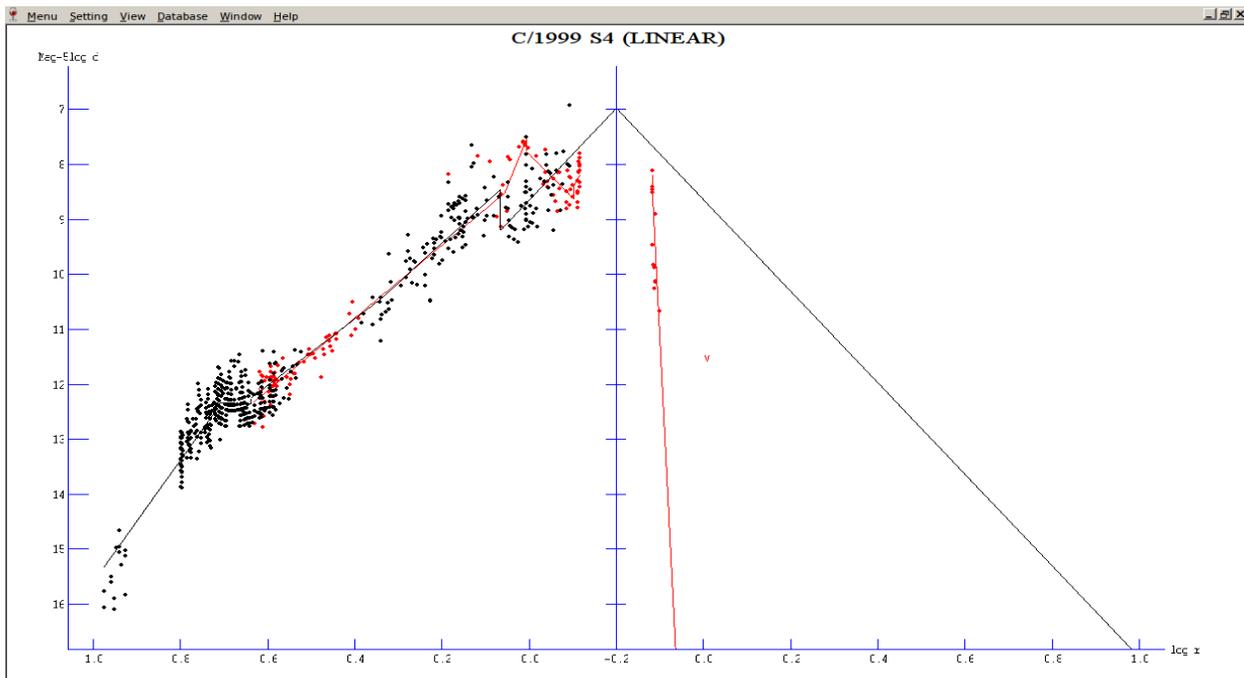

**Figure 11**. C/1999 S4 (LINEAR) show very similar evolution of its brightness compared to comet ISON, but much larger non-gravitational forces. Near Sun it exhibited few outbursts, which leads to its final disintegration.

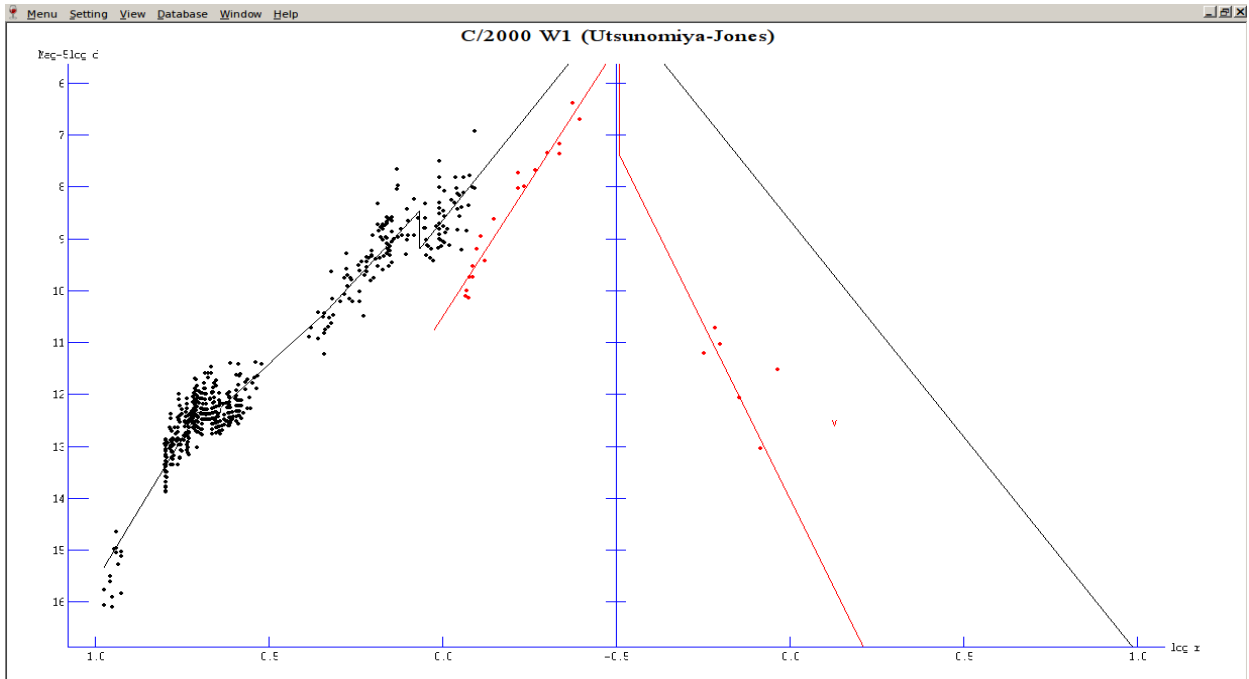

**Figure 12**. C/2000 W1 (Utsunomiya-Jones) was all time much fainter then ISON and observed only in short time.

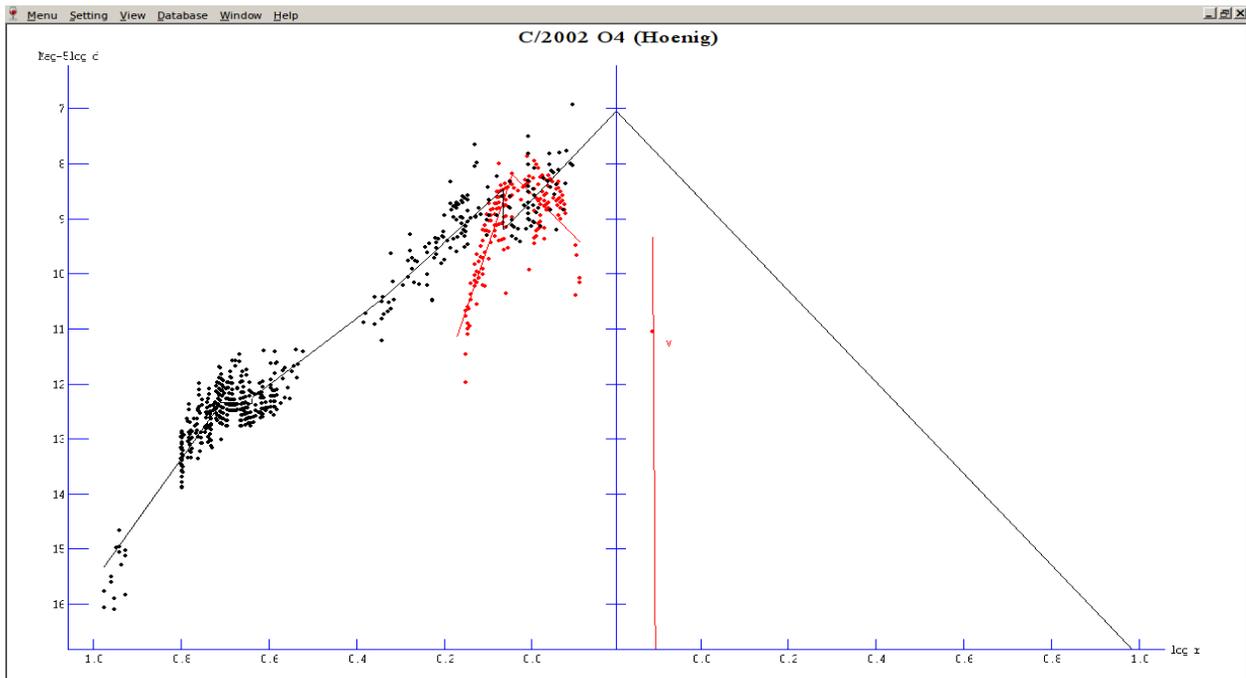

**Figure 13**. C/2002 O4 (Hoenig) was discovered after disintegration already take a place. During its most intensive phase it eventually shortly reached ISON activity level.

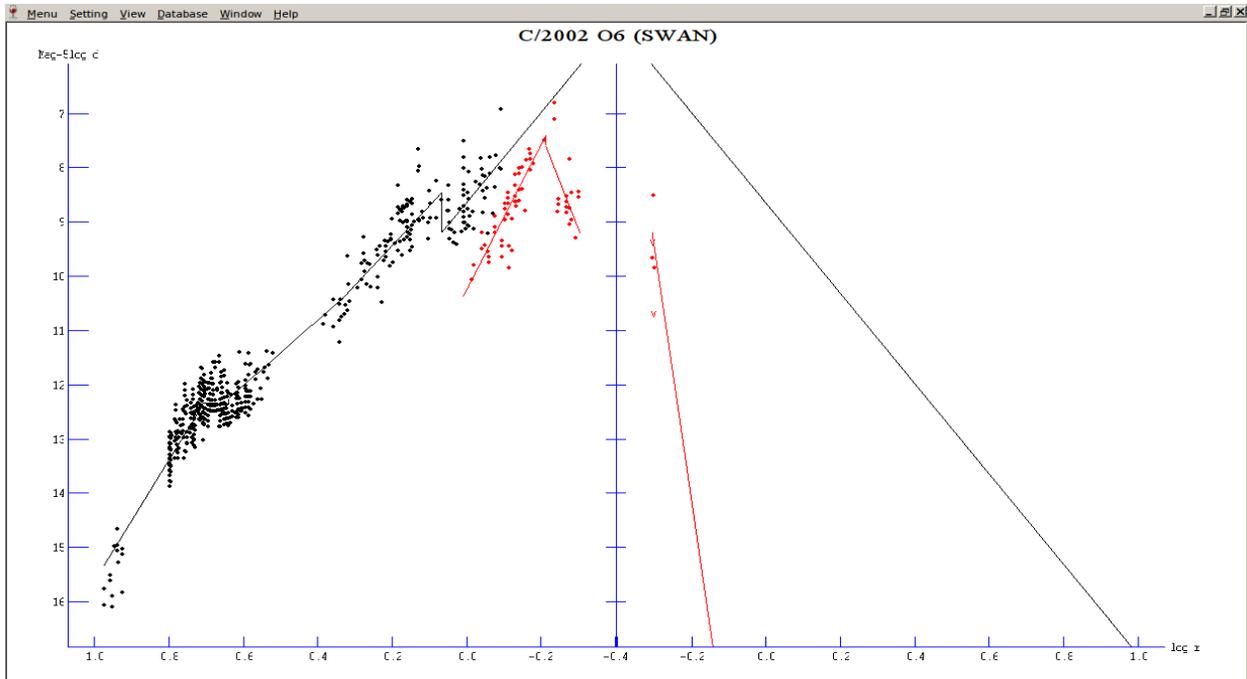

**Figure 14**. C/2002 O6 (SWAN) like many others, was probably discovered and observed already during its disintegration process.

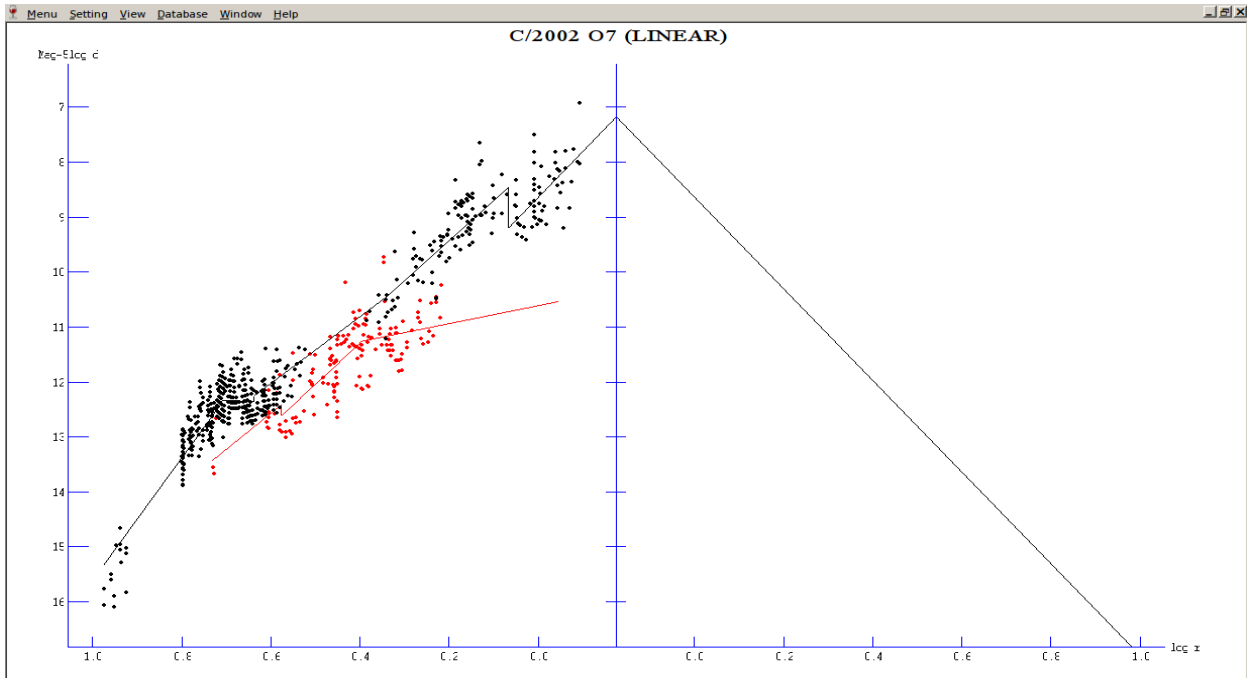

**Figure 15**. C/2002 O7 (LINEAR) exhibited very unique case of disintegration in very large distance from Sun.

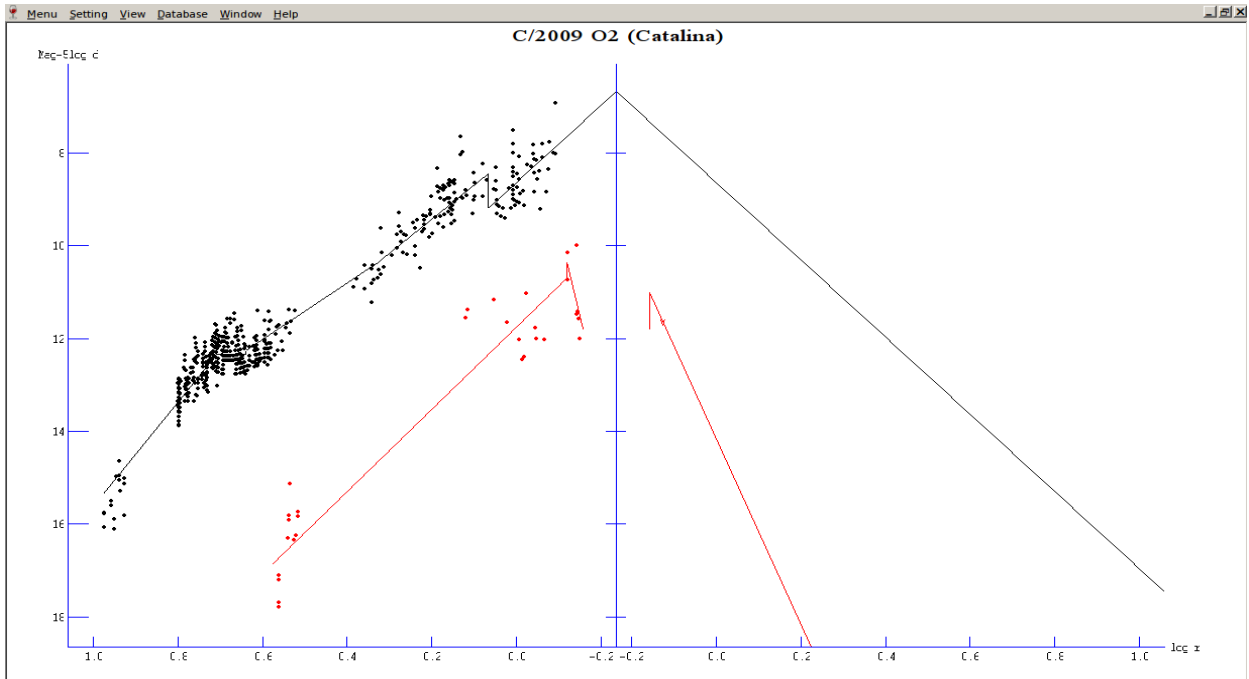

**Figure 16**. C/2009 O2 (Catalina) was extremely faint and poorly observed comet.

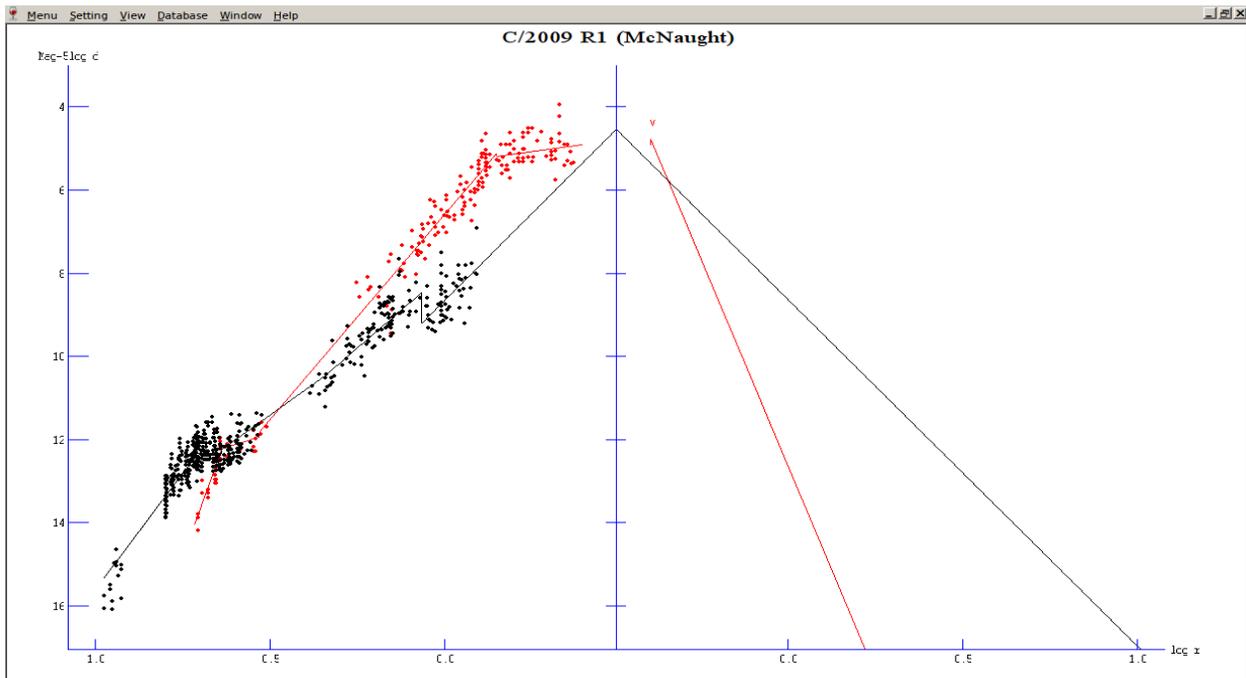

**Figure 17**. C/2009 R1 (McNaught) is one of most active comets amongst this sample, its fate is unknown and it is not proven that it really disintegrated.

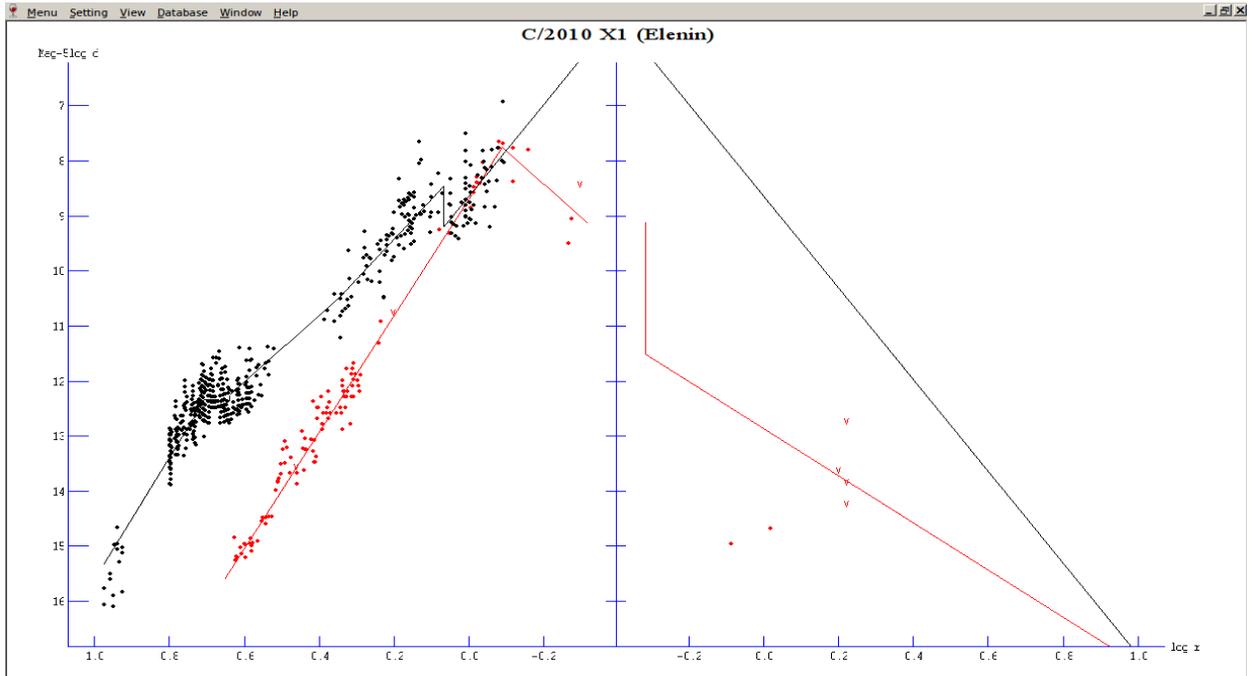

**Figure 18**. C/2010 X1 (Elenin) despite its original very small activity exhibit very strong and continuous brightening terminated with very fast disintegration, before which it shortly reached actual level of comet ISON activity.

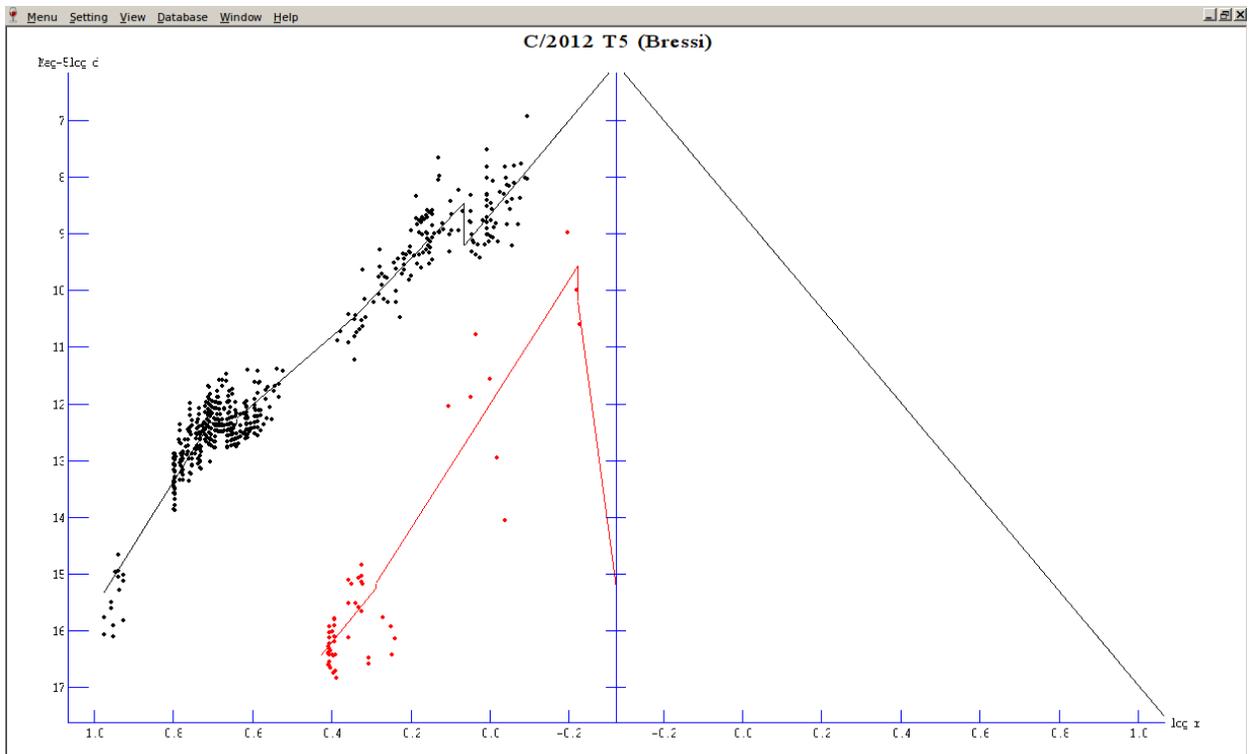

**Figure 19**. C/2012 T5 (Bressi) could be last example of extinct comet, unfortunately its light curve is poorly known and at least 2 outburst occurred before it disappeared, its demise is not confirmed but very probable.